\documentclass[a4paper,fleqn,usenatbib]{mnras}
\usepackage{newtxtext,newtxmath}
\usepackage[T1]{fontenc}
\usepackage{ae,aecompl}

\usepackage{graphicx}	
\usepackage{amsmath}	
\usepackage{amssymb}	
\usepackage[
]{hyperref}
\usepackage[usenames,dvipsnames]{xcolor}
\usepackage{url}
\usepackage{multirow}

\newcommand{\kms}{{\,\mathrm{km\ s^{-1}}}}
\newcommand{\Msun}{{\,\mathrm{M}_\odot}}

\newcommand{\masyr}{\,\mathrm{mas}\,\mathrm{yr}^{-1}}

\DeclareRobustCommand{\Figref}[1]{Fig.~\ref{#1}}
\DeclareRobustCommand{\Tabref}[1]{Table~\ref{#1}}
\DeclareRobustCommand{\Secref}[1]{Sec.~\ref{#1}}

\title[Space astrometry of the very massive
$\sim$$150\,M_\odot$ candidate runaway star VFTS682]{\emph{Gaia} and
  HST astrometry of the very massive $\sim$$150\,M_\odot$  candidate runaway star VFTS682}

\author[Renzo et al.]{M.~Renzo$^{1}$
  , S.~E.~de~Mink$^{1}$, D.~J.~Lennon$^{2,3}$, I.~Platais$^{4}$,
  R.~P.~van~der~Marel$^{4,5}$,
  \newauthor{E.~Laplace$^{1}$, J.~M.~Bestenlehner$^{6}$, C.~J.~Evans$^{7}$,
    V.~H\'enault-Brunet$^{8}$,  S.~Justham$^{9,10,1}$,
  }
\newauthor{A.~de~Koter$^{1,11}$,
      N.~Langer$^{12}$, F. Najarro$^{13}$, F.~R.~N.~Schneider$^{14}$, J.~S.~Vink$^{15}$}}

\date{Accepted 2018 October 12. Received 2018 October 4; in original
  form 2018 August 28 \\ Affiliations
can be found at the end of this manuscript}

\pubyear{2018}

\begin{document}
\label{firsxtpage}
\pagerange{\pageref{firstpage}--\pageref{lastpage}}
\maketitle

\begin{abstract}
 
How very massive stars form is still an open question in
astrophysics. VFTS682 is among the most massive stars known,
with an inferred initial mass of $\gtrsim$$150\,M_\odot$. It is located
in 30 Doradus at a projected distance of 29\,pc from the central
cluster R136. 
Its apparent isolation led to two hypotheses: 
either it formed in relative isolation 
or it was ejected dynamically from the cluster. 
We investigate the kinematics of
VFTS682 as obtained by \emph{Gaia} and \emph{Hubble Space Telescope} astrometry. We derive a projected velocity relative to
the cluster of $38\pm17\kms$ ($1\sigma$ confidence interval). Although
the error bars are substantial, two independent measures
suggest that VFTS682 is a runaway
ejected from the central cluster. This hypothesis is further supported by a variety of
circumstantial clues. The central cluster is known to harbor 
other stars more massive than $150\,M_\odot$ of similar spectral
type and recent astrometric studies on VFTS16 and VFTS72
provide direct evidence that the cluster can eject some of its most
massive members, in agreement with theoretical predictions.
If future data confirm the runaway nature, this would make VFTS682
the most massive runaway star known to date. 
\end{abstract}

\vspace{-30pt}
\section{Introduction}
\label{sec:intro}

How massive stars form is one of the major longstanding questions in astrophysics
\citep[e.g.,][]{zinnecker:07}. 
%
Obtaining clues from observations is challenging, because massive stars are intrinsically rare, 
evolve fast, typically reside in dense groups, and remain enshrouded in
their parent cloud during the entirety of the formation
process. Important progress has been made on the theoretical side,
\citep[e.g.][]{bate:09,kuiper:15,rosen:16}, but the simulations 
remain challenging.  

It has been proposed that most, if not all, stars form in clusters
\citep[and references therein]{lada:03}. In this picture, field stars are primarily the result
of the dissolution of dense groups.  However, a small but significant population
of massive stars exists in relative isolation, far from dense clusters
or OB associations and their origin remains a matter of debate
\citep{gvaramadze:12, lamb:16,ward:18}. One hypothesis to explain
the population of relatively isolated massive stars is that they
formed in the field \citep[e.g.,][]{parker:07}. Another
hypothesis is that these massive stars were ejected from the clusters
in which they formed. Such ejections may result from dynamical interactions \citep[e.g.,][]{poveda:67} or from the disruption of binary systems at the death of the companion  star \citep[e.g.,][]{blaauw:61, renzo:18}. 

\vspace*{-9pt}
One of the most extreme examples that has been considered in this
debate is the very massive star VFTS682  \citep[][]{bestenlehner:11,
  bressert:12}. This star is located in the field of the 30 Doradus
(30Dor) region in the Large Magellanic Cloud (LMC) and was studied as part of the multi-epoch spectroscopic VLT-FLAMES Tarantula Survey \citep[VFTS,][]{evans:11}. It is a hydrogen-rich Wolf-Rayet star of spectral type WNh5. Spectral analysis and comparison with evolutionary models lead to an inferred present-day mass of $137.8^{+27.5}_{-15.9}\,M_\odot$ corresponding to an initial mass of $150.0^{+28.7}_{-17.4}\,M_\odot$
\citep{schneider:18}. 
This makes VFTS682 one of the most massive stars known and one of the most extreme objects in the region.
From the spectral point of view, it is reminiscent of the very
massive stars in the core of the R136 cluster \citep{dekoter:97,crowther:10, crowther:16}. 
In particular, a remarkable similarity exists between the
spectra of VFTS682 and R136a3 \citep{rubio-diez:17}.

VFTS682 stands out by its relative isolation at a projected distance of 119.4 arcseconds, corresponding to 
$29$\,pc, from  the star cluster R136. \citet{bestenlehner:11}
considered two possible explanations for the offset: either
the star formed in situ as an isolated massive star, or it was ejected from  R136. N-body simulations 
indicate that the dynamical ejection of very massive stars like VFTS682 is
expected \citep[e.g.][]{fujii:11, banerjee:12}. The capability of a
young cluster to eject a large number of (very) massive stars is supported by
the recent findings of proper motion studies \citep[e.g.,][]{lennon:18, drew:18}.

\citet{platais:15,platais:18} analyzed multi-epoch \emph{Hubble Space
  Telescope} (HST) photometry and identified 10 stars likely ejected
from R136. \citet{lennon:18} investigated the kinematics of  isolated
O-type stars in the region using the second \emph{Gaia} data release
\cite[DR2,][]{gaia:16,brown:18} and showed that the proper motion,
postion and direction of the $\sim$$100\Msun$ star VFTS16 is consistent
with a runaway origin from R136. They found a less clear case for
VFTS72, and in both cases some tension between the kinematic age of
these stars and their apparent age remains.


In this paper we present an analysis of the new kinematic
constraints for VFTS682 provided by \emph{Gaia} DR2 and constraints
from HST proper motions  by \citet{platais:18}.   We discuss the
implications of the hypothesis that VFTS682 is a runaway star
ejected from R136.

\section{Observations}
\label{sec:sample}

\begin{table}
  \begin{center}
    \caption{Stellar parameters of VFTS682.}
    \begin{tabular}{llc|c|c}
      \hline\hline
      &Parameter & Units & Value & Ref.\\[2pt]     
       \hline
     & present day mass  & $[M_\odot]$ & $137.8^{+27.5}_
                                           {-15.9}$ & (1)
                                                    \\[2pt]
      & initial mass& $[M_\odot]$ & $150.0^{+28.7}_{-17.4}$ & (1)
      \\[2pt]
      &age & $[\mathrm{Myr}]$ & $1.0\pm0.2$ & (1) \\[2pt]
      & mass loss rate & $\log_{10}(\dot{M}/[M_\odot \ \mathrm{yr}^{-1}])$ & $-4.1\pm0.2$ & (2)\\[2pt]
      \hline
      \label{tab:star_param}
    \end{tabular}
    \vspace*{-5pt}
    {\tiny The quoted uncertainties are statistical, and do not include systematic
      effects in the modeling.
      (1)~\cite{schneider:18}
      (2)~\cite{bestenlehner:11}}
  \end{center}
\end{table}

\begin{table}
  \begin{center}
    \caption{Kinematics of VFTS682.}
    \begin{tabular}{llc|c|c}
      \hline
      \hline
      &Parameter & Units & Value & Ref.\\
      \hline
      \multicolumn{5}{l}{\emph{Absolute position and position relative
      to R136}} \\
      \hline
         &$\mathrm{RA}_\mathrm{VFTS682}$&[degrees] & \phantom{-0}84.73136339876477 
                     & (1) \\        
               &$\mathrm{DEC}_\mathrm{VFTS682}$&[degrees] &
                                                            \phantom{0}-69.07411071794998
                     & (1)  \\    
                                                     
                        &$\mathrm{RA}_\mathrm{R136}$&[degrees] & \phantom{00}84.6750
                     &  (2) \\        
               &$     \mathrm{DEC}_\mathrm{R136}$&[degrees] &  \phantom{0}-69.1006
                     &  (2) \\       
        &$      \delta\mathrm{RA}$  &[mas] & \phantom{-00}0.0547                      
        &  (3, 5)
  \\        
               &$     \delta\mathrm{DEC}$  &[mas] & \phantom{-00}0.0268 
                     &  (3, 5) \\  
                       &$  d_\parallel$  & [arcsec]
                         & 119.4
                                 &  (3) \\
      &$L_\parallel$ & [pc] & 29 & (3) \\

                     \hline
           \multicolumn{5}{l}{\emph{Gaia absolute proper motion for VFTS682
      and the region}} \\
      \hline
          &$\mu_\mathrm{RA}$&[$\masyr$] & $1.84\pm 0.07$
                     & (1) \\        
               &$\mu_\mathrm{DEC}$&[$\masyr$] & $0.79\pm 0.08$
                     &  (1) \\        
                 & $\rho\,(\mu_\mathrm{RA}, \mu_\mathrm{DEC})$ &  & $0.0226$
                        & (1)  \\         
       &$\langle\mu_\mathrm{RA}\rangle_\mathrm{R136}$&[$\masyr$] & $1.74\pm0.01$
                        & (4) \\
      &$\langle\mu_\mathrm{DEC}\rangle_\mathrm{R136}$&[$\masyr$]
                & $0.70\pm0.02$ &  (4)\\
\hline


      \multicolumn{5}{l}{\emph{Gaia DR2 proper motion of VFTS682 relative
      to R136}}\\
      \hline
      &$\delta\mu_\mathrm{RA}$  &[$\masyr$] & $0.10\pm0.08$ & (1,6) \\
      &$\delta\mu_\mathrm{DEC}$  &[$\masyr$] & $0.08\pm0.10$ &  (1,6) \\
      &$\delta\mu_{Gaia}$  &[$\masyr$] & $0.13\pm0.09$ &  (1,6) \\
      &$v_\mathrm{2D}$  &[$\kms$] & $32\pm21$ & (1,6)\\  
      &$\theta_{Gaia}$  &[degrees] &  $14_{-31}^{+36}$  & (1,6)\\  
 %

 \hline     
      \multicolumn{5}{l}{\emph{HST proper motion of VFTS682 relative
      to R136}} \\
            \hline
      &$\delta\mu_\mathrm{RA, HST}$  &[$\masyr$] & $0.02\pm0.10$ & (5) \\
      &$\delta\mu_\mathrm{DEC, HST}$  &[$\masyr$] & $0.19\pm0.09$ &  (5) \\
       &$\delta\mu_\mathrm{HST}$  &[$\masyr$] & $0.19\pm0.09$ &  (5) \\
                  &$v_\mathrm{2D, HST}$  &[$\kms$] & $45\pm21$ & (5)\\  
                      &$\theta_\mathrm{HST}$  &[degrees] &   $-30_{-51}^{+24}$   & (1,6)\\  
      \hline
      \multicolumn{5}{l}{\emph{Weighted average relative proper motion
      for VFTS682}}\\
      \hline
      &$\delta\mu_\mathrm{RA, avg}$  &[$\masyr$] & $0.08\pm0.07$ & (6) \\
      &$\delta\mu_\mathrm{DEC, avg}$  &[$\masyr$] & $0.14\pm0.07$ &  (6) \\
       &$\delta\mu_\mathrm{avg}$  &[$\masyr$] & $0.16\pm0.07$ &  (6) \\
      &$v_\mathrm{2D, avg}$  &[$\kms$] & $38\pm17$ & (6)\\
      \hline
       \multicolumn{5}{l}{\emph{Expected proper motion if ejected from
      R136 at age zero}} \\
      \hline
      &$v_\mathrm{2D}$  &[$\kms$] & $29\pm 6$ & (3) \\  
      &$\theta$  &[degrees] &  $\sim0$  & \\ 
       \hline
    \end{tabular}
    {\tiny The error on the RA and DEC positions, are of order
      $\sim$$0.01\masyr$ in \emph{Gaia}
      DR2. Assuming a distance of 50\,kpc, $1\masyr$ corresponds to $237\kms$.
$\rho(\mu_\mathrm{RA},\mu_\mathrm{DEC})$ is the
      correlation coefficient. The position angle
      $\theta$ is defined such that $\theta=0$ for radial motion
      away from R136. We neglect the error bars on
      $\langle\mu_\mathrm{RA}\rangle_\mathrm{R136}$ and $\langle\mu_\mathrm{DEC}\rangle_\mathrm{R136}$ to determine the uncertainty
      on $\theta_{Gaia}$.
      (1)~\cite{brown:18},
      (2)~\cite{henault-brunet:12},
      (3)~\cite{bestenlehner:11},
      (4)~\cite{lennon:18}, 
      (5)~\cite{platais:18} and
      (6)~{\color{blue}this study}.
    }
  \end{center}
  \label{tab:vfts682}
\end{table}
The WNh5 star VFTS682, located at right ascension (RA)
05$^\mathrm{h}$38$^\mathrm{m}$55.510$^\mathrm{s}$  and declination
(DEC) \mbox{-69$^\mathrm{o}$04'26.72''} (J2000), was observed as part of the multi-epoch, spectroscopic VFTS campaign covering $\lambda$4000--7000 \citep[][]{evans:11}. 
\citet{bestenlehner:11}  analyzed the spectra to infer the stellar
parameters and measured a visual extinction of $A_V=4.45\pm0.12$, implying a
luminosity of $\log_{10}(L/L_\odot) =  6.5\pm0.2$, making this one of
the most luminous stars in the region. The absence of periodic radial
velocity (RV)
variations suggests that the star is unlikely to have close
companions \citep[][]{bestenlehner:11}, unless the orbital inclination is
very high. Bayesian fits of the stellar
parameters against evolutionary tracks \citep{brott:11, kohler:15}
using the BONNSAI code \citep{schneider:14,schneider:17} provide
estimates for the age, present mass and initial mass, 
see
\Tabref{tab:star_param}. 

VFTS682 is not a bright X-ray point source. It was not detected in the
\emph{Chandra} survey of \cite{townsley:06}, and shows a few counts in
the deeper survey of \cite{townsley:14}.
The X-ray luminosity of VFTS682
is significantly lower than known massive binaries in the region, which suggests the absence of
colliding winds. These would be expected in the presence of companions
even for extreme mass ratios, given the large mass of
VFTS682.

This star is also relatively isolated in the \mbox{(near-)infrared}. The nearest bright (near-)infrared sources detected by
\emph{Spitzer} \citep{meixner:06} and resolved in the \emph{VISTA}
Magellanic Clouds Survey \citep{cioni:11} are located at a distance of
about 10 arcsecond, i.e. about 2.4 pc. \cite{walborn:13} speculate
that these nearby young stars may represent a case of star formation triggered by the wind of VFTS682.

The V-band light curve of VFTS682  shows
variations at a $\sim$10\% level on a timescale of years, which is
unusual for Wolf-Rayet stars and more typical for Luminous Blue
Variable (LBV) stars \citep{udalski:08, bestenlehner:11}. The source also
shows a mid-infrared excess \citep{gruendl:09}.

Estimates of the radial velocity are complicated by the variable,
possibly inhomogeneous, optically thick wind  typical of emission line
stars.  We therefore caution
against over-interpreting the existing radial velocities estimates. \citet{bestenlehner:11} estimate a mass loss rate of
$\log_{10}(\dot{M}/[M_\odot \ \mathrm{yr}^{-1}])=-4.1\pm0.2$, not accounting for the
possible effect of clumping.  
They estimate a RV of
$300\pm10\kms$ using the  N{\footnotesize V}\,$\lambda4944$ line, which
is offset from the average radial velocity of the region of
$270\pm10\kms$. This was suggested as indicating a runaway nature, but
it is no proof of it. 
\cite{bressert:12} note an offset between the RV of
the star and the nebular lines from the gas filaments in its
vicinity. This is in line with the expectation that the star was not
formed in situ. Given these issues, we refrain from using the RV measurements in this
work, and focus on the velocities on the plane of the sky. 

We adopt a distance to the LMC of $50$\,kpc. The error on
the distance determination is small ($\lesssim2\%$,
\citealt{pietrzynski:13}) and any possible offset in the radial
direction between R136 or VFTS682 and the distance we adopted for the
LMC is probably much smaller ($\sim$$0.5\%$, e.g., \citealt{luks:92}). These uncertainties are negligible compared to the errors in the proper motion discussed below.  

\begin{figure}
  \centering
  \includegraphics[width=0.35\textwidth]{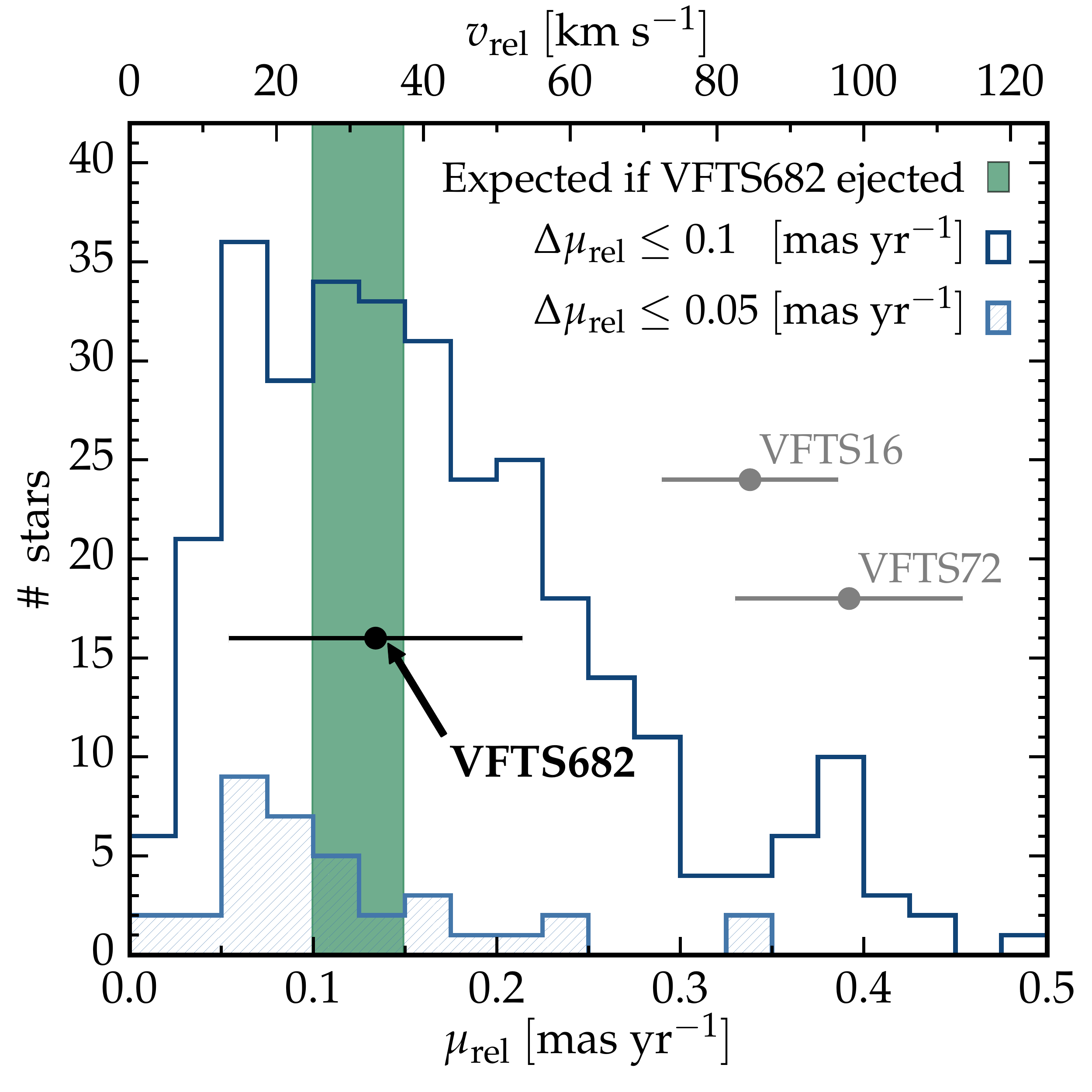}\\
  \vspace*{-5pt}
  \includegraphics[width=0.35\textwidth]{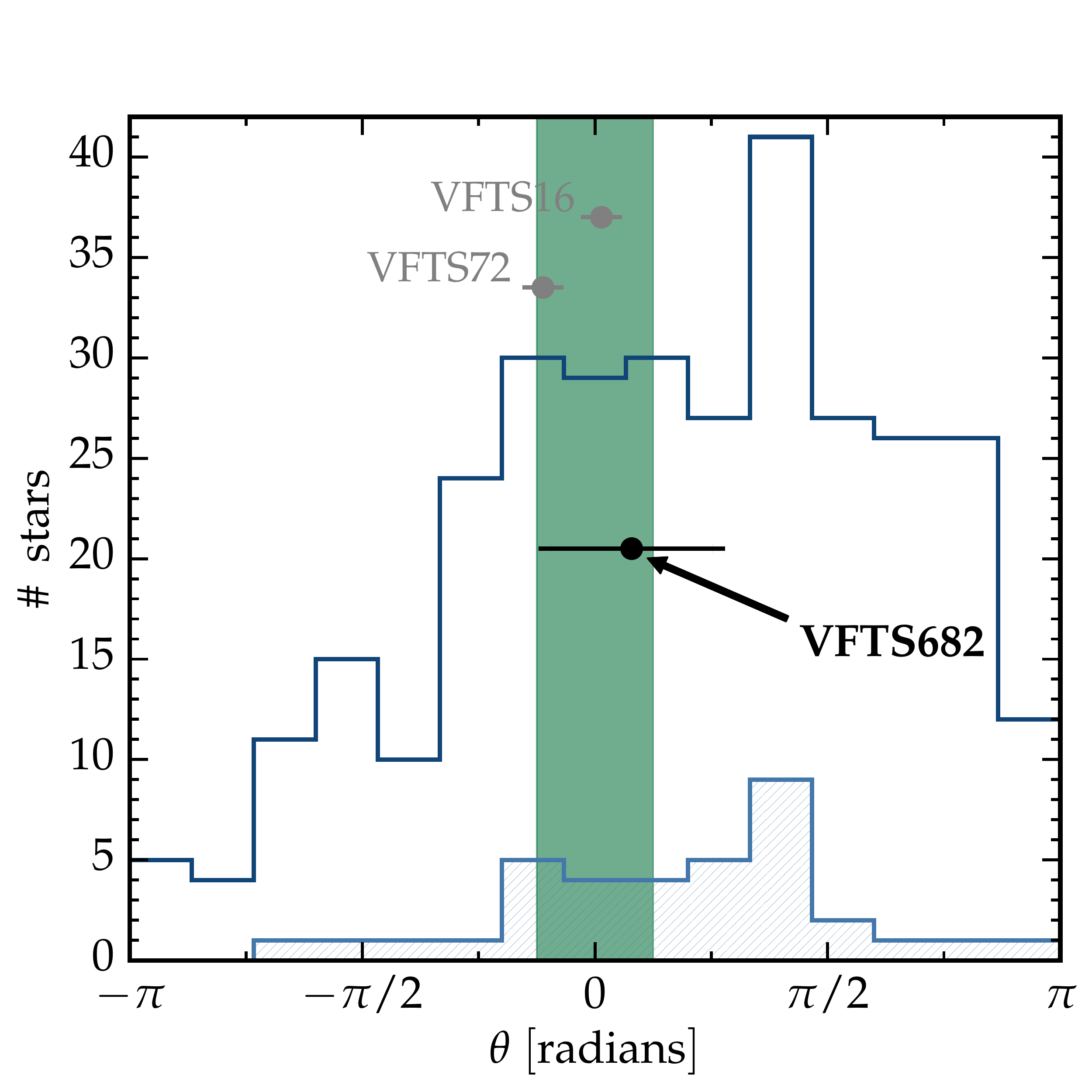}
  \vspace{-5pt}
  \caption{Distribution of OB-type and Wolf-Rayet stars in proper
    motion relative to R136 (top panel) and proper motion position angle
    (bottom panel), from \emph{Gaia} DR2. Although VFTS682 is not an outlier, its relative
    proper motion matches the value expected for an early dynamical
    ejection (see \Secref{sec:results}). In both
    panels, the dark blue histograms contain 317 
    stars with error bars smaller than $0.1\,\mathrm{mas \
      yr^{-1}}\simeq25\,\mathrm{km\ s^{-1}}$ at 50\,kpc and the
    lighter blue histograms contain 36 stars with error bars smaller than $0.05\,\mathrm{mas \
    yr^{-1}}$. The peak at $\theta\simeq\pi/2$ in the bottom panel is due to stars
  belonging to NGC2060.}
\vspace{-5pt}
  \label{fig:dist}
\end{figure}

\begin{figure*}
  \centering
  \includegraphics[width=\textwidth]{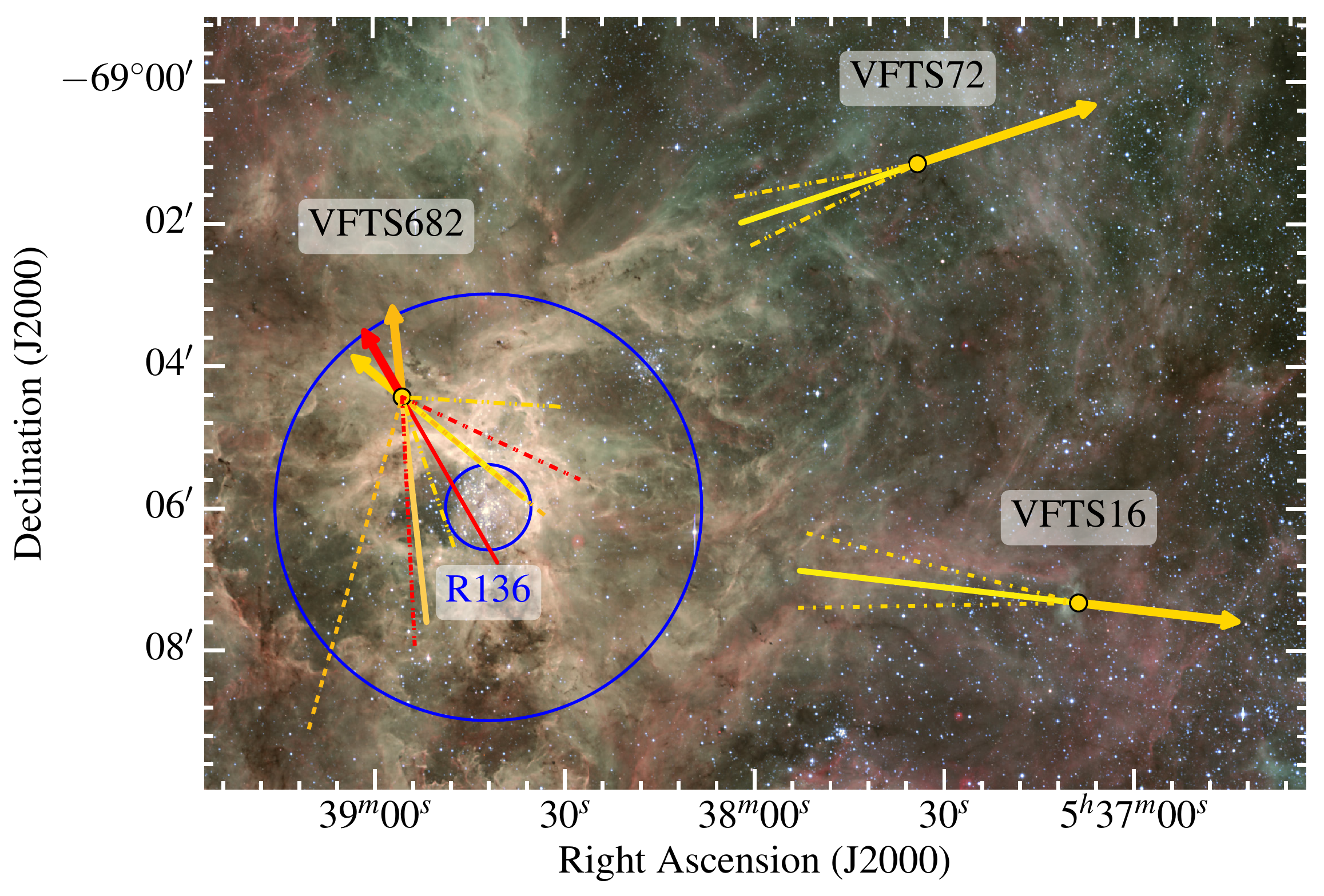}  
  \caption{The thick red arrow shows the proper motion
    relative to R136 for VFTS682 from averaging the \emph{Gaia} DR2 and HST
  astrometry, multiplied by 0.4\,Myr. The extension in the opposite
  direction is proportional to the apparent age of the star, and the
  thin lines illustrate the error cone on the potential origin. The
  yellow (orange) arrows show the \emph{Gaia} DR2 (HST) results alone. The two blue circles indicate the regions of radii 0.01 and 0.05
  degrees around the core of R136.}
\vspace{-5pt}
  \label{fig:main}
\end{figure*}

\subsection{ \emph{Gaia} astrometry for VFTS682\label{data:gaia}}

VFTS682 is identified with the source id 4657685637907503744 in the
\emph{Gaia} DR2 catalog 
  as a 15.65 mag star in the G band
\citep{gaia:16,brown:18}.   The number of visibility periods,
i.e., groups of observations separated from each other by at
least four days, used in the astrometric solution is seventeen. The reported astrometric excess noise is zero.  These values
suggest that the \emph{Gaia} DR2  data for VFTS682 are
reliable.

Gaia provides absolute proper motions.  To determine the
proper motion relative to R136, we follow  \citet{lennon:18} to
define the motion of the local frame of reference using the average
proper motion of nearby stars with reliable astrometric data (see \Tabref{tab:vfts682}).  They
selected bright ($G<17$) stars within 0.05 degrees of R136 and exclude
sources with proper motion error bars greater than $0.1\masyr$ in both
coordinates (see their  Sect.~2.1).  Using this definition of the
local frame, we compute the relative proper motion
$\delta\mu_\mathrm{RA}$ and $\delta\mu_\mathrm{DEC}$.   We also compute the total projected 2D velocity
($v_\mathrm{2D}$) and the angle $\theta$ between the direction of motion and the vector
connecting the center of R136 with the current position of VFTS682.
All kinematic quantities are provided in
\Tabref{tab:vfts682}. 

\vspace*{-15pt}
\subsection{HST (WFC3/UVIS) astrometry for VFTS682}

The 30 Dor region was targeted by a two-epoch photometric campaign
with HST providing observations in the F775W filter in October 2011
and October 2014 (GO-12499; P.I.: D.~J.~Lennon). 
\citet{platais:15, platais:18} analyzed the HST data to determine the
relative proper motions and identify candidate runaway stars. The
brightest stars (V$<$14) are saturated in the data set and have been
excluded from the analysis. The high extinction around VFTS682
makes it redder and fainter \citep[$V=16.08$,
$B-V=0.58$,][]{evans:11}, hence it has reasonably accurate HST astrometry
with the WFC3/UVIS camera. This star did not pass a full set of stringent
conditions to be considered as a candidate runaway \citep[][]{platais:18}.
In retrospect, VFTS682 may have been included in the list of likely
OB runaway stars, and it is identified
with the ID source 330375 in their catalog. 
Therefore, their measurements provide a useful
complementary estimate of the proper motion of VFTS682 which is independent from the \emph{Gaia} data. 
 
The HST study provides proper motions that are relative to the bulk motion
of the majority of the stars in the field of view. The full 30 Dor
field is covered by different pointings and there is some systematic
distortion.  However, even for stars far from 30 Dor the effect is
small, no more than $0.05\masyr$ across the whole 30 Dor field. The
effect is much smaller for stars close to the center of the field, such as
VFTS 682 \citep{platais:18}.  We can therefore use the relative proper
motion (\Tabref{tab:vfts682}) as a good estimate for the proper motion relative to R136. 
 
\section{The kinematics of VFTS682}
\label{sec:results}

The black points in~\Figref{fig:dist} show the proper motion 
(top panel) and the projected flight direction (bottom panel) relative
to R136 of
VFTS682 from \emph{Gaia} DR2. 
Dynamical ejections from the
cluster should 
produce close to radial ejections, i.e.~$\theta\simeq0$. The green
vertical bands highlight the expectations for these two quantities. 
The width in the top panel is determined
by the error bars on the star's apparent age, and we assume a width of
$45$ degrees in the bottom panel. For comparison, we also show the relative
proper motion of VFTS16 and VFTS72 (gray points), and the distribution
in relative proper motion and flight direction for all the VFTS OB-type and Wolf-Rayet stars with
\emph{Gaia} DR2 errors on the proper motion
components of less than $0.1\,\mathrm{mas\ yr^{-1}}$ (dark blue lines,
including VFTS682), and less than $0.05\,\mathrm{mas\ yr^{-1}}$ (light hatched
blue). Although the error bars are substantial and VFTS682 is not
an outlier compared to other OB-type and Wolf-Rayet stars, the agreement
suggests that the star is indeed a runaway as suggested by
\cite{bestenlehner:11}. 

Subtracting
the mean motion of R136, we obtain relative proper motions (projected
velocities) of $\delta \mu_{Gaia}=0.13\pm 0.09\masyr (32\pm 21\kms)$ and
$\delta \mu_{HST}=0.19\pm 0.09\masyr (45 \pm 21\kms )$. Both values
are consistent with each other, but also with no motion relative to R136
within $2\sigma$. The average (weighted with $1/\sigma^2$) of these two independent
measurements is $\delta \mu_\mathrm{avg}=0.16\pm0.07\masyr
(38\pm17\kms)$. 

Figure~\ref{fig:main} shows the motion of VFTS682 relative to R136
projected on the
sky. We also show VFTS16 and VFTS72 
\citep[see][]{lennon:18}. The yellow arrows are proportional to
the relative proper motion from \emph{Gaia} DR2, the orange
line illustrates the relative proper motion from HST, and the 
red arrow shows the averaged result. The error cone
on the direction of motion is illustrated by the corresponding
extension in the direction opposite to the motion, and we also show
the most likely origin of the stars accounting for their apparent age
($0.7\pm0.1$\,Myr and $0.4^{+0.8}_{-0.4}$\,Myr for VFTS16 and VFTS72,
respectively \citealt{schneider:18}).
This figure illustrates that R136 is the most likely origin of these stars, although the large error bars
prevent a robust identification for VFTS682, and there is some tension
between the apparent age and the present day distance from the cluster
core for VFTS16 and VFTS72 \citep[][]{lennon:18}. We note that
VFTS72 has a small radial velocity, while VFTS16 (and possibly VFTS682) has
a large peculiar radial velocity, and therefore accurate distances along
the line of sight are
needed to constrain the flight direction in three dimensions.

Assuming VFTS682 indeed originates from R136, we can calculate its kinematic
age as:
\begin{equation}
  \label{eq:kin_age}
  \tau_\mathrm{kin} = \frac{d_\parallel}{\delta\mu_\mathrm{avg}} \simeq
  \frac{119.4\,\mathrm{arcsec}}{0.16\masyr} \simeq 0.7\pm\,0.3\, \mathrm{Myr} \ \ ,
\end{equation}
where $d_\parallel = 119.4\,\mathrm{arcsec}$ is the angular distance from VFTS682 to
the core of the cluster \citep[][]{bestenlehner:11}. 
The kinematic age $\tau_\mathrm{kin}$ is consistent with an early
ejection from the cluster (see \Tabref{tab:star_param}).

In summary, both \emph{Gaia} and HST relative proper motions are consistent with the dynamical ejection of
VFTS682 from the cluster, although we cannot confidently rule out the
hypothesis of in situ formation. 
\vspace*{-20pt}
\section{Discussion}
\label{sec:discussion}

Based on our results, we consider that VFTS682 is
potentially the most massive
runaway known to date, with a two-dimensional
projected velocity with respect to R136 of
$38\pm17\kms$ (taking a weighted average of \emph{Gaia} DR2 and HST
results). Due to the large error bars, this result will need
to be revisited with future astrometric data. 

If confirmed, 
isolated star formation is
\emph{not} required to explain the isolation of VFTS682. Its proper motion suggests that it was ejected from the cluster R136
$0.7\pm0.3$\,Myr ago, which is compatible with the evolutionary age of
the star. If the
cluster age \citep[$\lesssim2$\,Myr,][]{crowther:10, sabbi:12} is
indeed smaller than the shortest stellar lifetime
\citep[$\sim$3\,Myr,][]{brott:11,kohler:15, zapartas:17}, the ejection of VFTS682
from the disruption of a massive binary by a supernova is excluded. 
The kinematic age we infer is smaller than the kinematic age of
$\sim$1.5\,Myr for VFTS16 found in \cite{lennon:18}, which indicates
that VFTS682 was ejected later than VFTS16, and potentially later than
VFTS72 too.

If the star were ejected dynamically, its isolation makes it an ideal target to constrain the stellar physics of
stars with masses well above $\sim$$100\,M_\odot$ in the inner
cluster, while avoiding
crowding issues. Moreover, 
 its exceptionally large mass raises the question of which stars must populate
the core of the cluster. N-body dynamics typically ejects the least
massive star among those interacting \cite[although the
  dynamical ejection fraction increases with mass because of mass
  segregation, e.g.,][]{banerjee:12}. Just
based on the kinematic properties of VFTS682, we would expect several
stars with initial masses larger than $\sim$$150\,M_\odot$ in the
cluster R136, as it is observed.

The N-body simulations of \citet{banerjee:12} suggest that VFTS682 was ejected from R136. They
demonstrated that the cluster potential does not significantly change
the velocity of the star after the ejection.
To eject such a massive object, the cluster is
expected to have produced a large number of massive runaways,
and their simulation suggest a significant incidence of
 (dynamically driven) stellar mergers both in the cluster and among
 the stars ejected. Indeed, several 
isolated massive stars are observed in the region \citep[][]{evans:10,lennon:18}, some with known
large radial velocities and/or proper motion. 
A comprehensive study of the kinematic
properties of all the massive stars surrounding R136 could shed light
on whether some can be unequivocally identified as merger products,
but also on the initial conditions for the cluster dynamics
\citep[e.g.,][]{oh:16}, and whether it formed via a monolithic collapse, or
as a (potentially ongoing) merger of several sub-structures \citep[e.g.,][]{sabbi:12}.

Also \cite{fujii:11} suggest that
early in the evolution of a cluster, dynamical interactions form an extremely
massive binary, which then tightens its orbit by ejecting other
stars. The spectral similarities between VFTS682 and stars in the core
of R136 are in agreement with this ``bully binary'' model. Interpreting the kinematics of VFTS682 through the lens of their simulations
suggests the presence of a close binary with total mass
$M_1+M_2\gtrsim 300\,M_\odot$ in the core of the cluster. The
difference between the cluster age and the kinematic age of VFTS682 puts an
upper limit to the timescale to form the ``bully binary'' in
R136 of $\sim$$1.3\,\mathrm{Myr}$. 
Such a binary 
might be a 
candidate for a dynamically formed progenitor system of
a binary black-hole, provided that stars this massive can avoid a
pair-instability supernova \cite[e.g.,][]{rakavy:67} at LMC
metallicity \citep[see also][]{langer:07, woosley:17}.
Similarly, the final fate of VFTS682 could be either a
pair-instability supernova without compact remnant formation, or collapse to a black hole. The amount of mass loss of these stars will determine their final core
mass and thus their final fate \cite[e.g.,][]{vink:15}.

VFTS682 is potentially the most massive runaway known to date, and its ejection
from the cluster R136 likely implies that it is only the ``tip of the
iceberg'' of massive runaways in the
region. Studies of this population, enabled by recent and future 
observations will put constraints on the evolution
of these stars, together with the formation and evolution of
the central cluster itself.\\
  
  { \small
    We thank P.~Crowther, J.~Heyl, M.\,C.\,Ramirez-Tannus, S.~N.~Shore,
    and S.\,Torres, and the HSTPROMO Collaboration
    for help and discussions. SdM acknowledges
    the European Unions Horizon 2020 research and innovation programme
    from the European Research Council (ERC), Grant agreement No. 715063
    and VHB the NRC-Canada Plaskett Fellowship. This study has used data from the ESA mission {\it Gaia} (\url{https://www.cosmos.esa.int/gaia}), processed by 
    DPAC, \url{https://www.cosmos.esa.int/web/gaia/dpac/consortium}.
  }


\label{lastpage}

\section*{Affiliations}
\noindent $^{1}${Astronomical Institute Anton Pannekoek, University of
  Amsterdam, 1098 XH Amsterdam, The Netherlands}

\noindent $^{2}$ {ESA, European Space Astronomy Centre, Apdo. de Correos 78,
    E-28691 Villanueva de la Ca\~nada, Madrid, Spain} 

  \noindent $^{3}$ {Instituto de Astrof\'isica de Canarias, E-38205 La Laguna, Tenerife, Spain}

  \noindent $^4$ {Department of Physics \& Astronomy, Johns Hopkins University, Baltimore, MD 21218, USA}

  \noindent $^{5}$ {Space Telescope Science Institute, 3700 San Martin Drive,
    Baltimore, MD 21218, USA}

  \noindent $^{6}${Department of Physics and Astronomy, Hicks Building,
    Hounsfield Road, University of Sheffield, Sheffield S3 7RH, UK}

  \noindent $^{7}${UK Astronomy Technology Centre, Royal Observatory Edinburgh, Blackford Hill, Edinburgh, EH9 3HJ, UK}

  \noindent $^{8}$ {National Research Council, Herzberg Astronomy \&
    Astrophysics, 5071 West Saanich Road, Victoria, BC, V9E 2E7,
    Canada}

  \noindent $^{9}$ {School of Astronomy \& Space Science, University of the Chinese
    Academy of Sciences, Beijing 100012, China}

  \noindent $^{10}$ {National Astronomical Observatories, Chinese Academy of
    Sciences, Beijing 100012, China}

  \noindent $^{11}$ {Institute of Astronomy, KU Leuven, Celestijnenlaan 200D, 3001 Leuven, Belgium}
  
  \noindent $^{12}$ {Argelander-Instit\"ut f\"ur Astronomie, Universit\"at Bonn,
    Auf dem H\"ugel 71, 53121, Bonn, Germany}

  \noindent $^{13}$ {Centro de Astrobiolog\'ia, CSIC-INTA, Carretera de Torrej\'on a Ajalvir km-4, E-28850 Torrej\'on de Ardoz, Madrid, Spain}

  \noindent $^{14}$ {Department of Physics, University of Oxford, Keble Road,
    Oxford OX13RH, UK} 

  \noindent $^{15}$ {Armagh Observatory, College Hill, Armagh BT619DG, UK}

\end{document}